\documentclass[11pt]{article}
\usepackage{amsmath,amssymb,color,graphics,epsfig,cite}

\usepackage{amsfonts}

\newcommand{\hoch}[1]{$\, ^{#1}$}

\newcommand{\be}{\begin{equation}}
\newcommand{\ee}{\end{equation}}
\newcommand{\bea}{\setlength\arraycolsep{2pt} \begin{eqnarray}}
\newcommand{\eea}{\end{eqnarray}}
\newcommand{\nn}{\nonumber}

\def\ft#1#2{{\textstyle{\frac{\scriptstyle #1}{\scriptstyle #2} } }}
\def\fft#1#2{{\frac{#1}{#2}}}

\def\0{{\sst{(0)}}}
\def\1{{\sst{(1)}}}
\def\2{{\sst{(2)}}}
\def\3{{\sst{(3)}}}
\def\4{{\sst{(4)}}}
\def\5{{\sst{(5)}}}
\def\6{{\sst{(6)}}}
\def\7{{\sst{(7)}}}
\def\8{{\sst{(8)}}}
\def\sst#1{{\scriptscriptstyle #1}}

\begin{document}

\begin{center}
{\Large {\bf Superradiant Instability of Charged Extremal Black Holes in Einstein-Born-Infeld Gravity }}

\vspace{20pt}

{\large Zhe-Hua Wu\hoch{1} and H. L\"u\hoch{1,2}}

\vspace{10pt}

{\it \hoch{1}Center for Joint Quantum Studies and Department of Physics,\\
School of Science, Tianjin University, Tianjin 300350, China }

\bigskip

{\it \hoch{2}Joint School of National University of Singapore and Tianjin University,\\
International Campus of Tianjin University, Binhai New City, Fuzhou 350207, China}

\vspace{40pt}

\underline{ABSTRACT}
\end{center}

We study charged scalar perturbations of charged extremal black holes in Einstein-Born-Infeld theory. Our numerical results indicate that these black holes all suffer from superradiant instability by the unstable quasi-bound states, regardless how small the coupling constant is. We therefore provide a new example that the superradiant stability of the Reissner-Nordstr\"om black hole is a fine-tuned result, as in the case when it is embedded in the STU supergravity model. The work is also motivated by the weak gravity conjecture since at the linear coupling constant level, the theory belongs to a subsect of four-derivative corrections in the effective field theory. Our results appear to support the notion that the black holes do decay when gravity is weaker by the correction, but the decaying halftime requires nonlinear effects and cannot be seen at the level of linear coupling constant. The full nonlinear effects also indicate that the black holes can decay even when gravity is stronger.

\vfill{\footnotesize wuzh\_2022@tju.edu.cn\ \ \ mrhonglu@gmail.com}

\thispagestyle{empty}
\pagebreak

\tableofcontents
\addtocontents{toc}{\protect\setcounter{tocdepth}{2}}

\newpage

\section{Introduction}

Universal gravity and electrostatic forces are two known long-range forces in nature. They can be united classically in Einstein-Maxwell (EM) gravity, which predicts the existence of Reissner-Nordstr\"om (RN) charged black holes. Despite of the fact that it is unlikely for them to exist in nature, RN black holes, especially in the extremal and near-extremal limits, remain one of the favourite spacetime geometry for theoretic study. The perfect balance between gravity and the electrostatic force is preserved in the extremal RN black holes under the no force condition, where mass and charges are exactly the same. EM gravity can be naturally embedded in supergravity, the low-energy effective theory of strings. In this framework, the Maxwell field is not fundamental, but composite of appropriate components such as D-branes and Kaluza-Klein monopoles, which can be described in STU supergravity model \cite{Duff:1995sm}. This stringy framework provides the first example of counting the microscopic entropy for (extremal) black holes \cite{Strominger:1996sh}. The RN black hole also provides an excellent toy model to test the weak gravity conjecture (WGC), which states that gravity should be weaker than any of gauge interactions \cite{Arkani-Hamed:2006emk}. The no-force condition of the extremal RN black hole is therefore expected to be broken under quantum effect so that gravity becomes weaker than the electric force. This was indeed confirmed in the context of effective field theory under the leading order correction \cite{Cheung:2018cwt}.  By contrast, gravity becomes stronger than the centrifugal repulsion under quantum correction in this setup of effective field theory \cite{Ma:2020xwi}. The WGC suggests that the zero-temperature particle-like extremal RN black hole, which can be large, will decay into smaller ingredients under quantum correction. It is thus natural to ask what could be the concrete mechanism for such instability.

Charged black hole has an intriguing phenomenon called superradiant effect, see e.g.~\cite{Press:1972zz,Bekenstein:1973mi,Teukolsky:1974yv,Starobinsky:1973aij}. One can extract energy and charges from a charged black hole by throwing in a charged matter fields such as a scalar, even though one imposes only the ingoing boundary condition on the black hole horizon. This is analogous to the Penrose process and leads to a nature question: will this superradiant effect cause instability of the black hole? Indeed such instability was demonstrated in rotating black holes \cite{Detweiler:1980uk,Dolan:2007mj}. However, It was convincingly argued in \cite{Hod:2012wmy} that the extremal RN black hole does not suffer from such instability under the charged scalar perturbation. This result was later generalized to non-extremal RN black holes \cite{Huang:2015jza,Hod:2015hza}, which should be expected since gravity is stronger.  This issue was later analysed \cite{Mai:2021yny,Mai:2022thu} in the context of STU supergravity model, which involves four different Maxwell field strengths, associated with different D-branes and Kaluza-Klein vectors. The RN black hole emerges as a fine-tuned object when all the four charges are exactly the same. It was shown that the charged black holes suffered from superradiant instability as long as not all the charges are exactly equal, no matter how small their difference is \cite{Mai:2022thu}. This shows that the superradiant stability of the RN black hole is a fine-tuning result of the general four-charge black holes in the supergravity model.

In this paper, we shall study the nonlinear generalization of the Maxwell theory. The best known example is the Born-Infeld (BI) theory. The analytic solution of charged black hole in Einstein-Born-Infeld (EBI) theory has long been know \cite{Born:1934gh,Hoffmann:1935ty}. (See also e.g.~\cite{Fernando:2003tz,Dey:2004yt, Cai:2004eh,Li:2016nll} for solutions with a cosmological constant.) We study the minimally coupled charged scalar perturbation of the extremal black hole. Our motivation is twofold. One is to examine the superradiant stability of the RN black hole in the context of a bigger theory, as this was done in the STU supergravity model. The other is inspired by the WGC, since if we take Taylor expansion of the BI coupling constant, the terms associated with the linear-order of the coupling constant belong to a subset of the general four-derivative effective field theory that describes the correction to the EM theory \cite{Cheung:2018cwt}. For the positive coupling, the perturbation satisfies the WGC requirement, namely \cite{Cheung:2018cwt}
\be
\Delta S>0\,,\qquad \Delta M<0\,.
\ee
Therefore, the extremal charged black holes in EBI gravity should be unstable no matter how small the coupling constant is.

Our conclusion is that extremal charged black holes in EBI are all superradiantly unstable, indicating again that the superradiant stability of the RN black hole is a fine-tuning result in the context of EBI theory. However, the detailed results make the question subtle to answer whether this instability is relevant to the WGC. Furthermore, the decay rate is extremely small compared to previous examples of unstable QBS's in literature.

The paper is organized as follows. In section 2, we review the EBI theory and its charged black hole. We illustrate that the theory can be used as a toy model for testing GWC. We consider charged scalar perturbations with the aim of constructing superradiantly unstable quasi-bound states (QBS's). We derive some necessary conditions for such states.  In section 3, we describe the two numerical methods to construct unstable QBS's. One is the shooting-target method and the other is the Chebyshev spectral method.  The reason for using two methods to do crosscheck is that the imaginary part of the complex frequency of the QBS's os exceedingly small. In section 4, we present numerical results that indicate the extremal charged black holes in EBI gravity are all superradiantly unstable.  We conclude the paper in section 5.

\section{Charged black hole and charged scalar perturbation}
\label{sec:theory}

In this section, we consider EBI gravity in four dimensions. We focus on electrically charged black holes. The relevant part of the Lagrangian is
\be
{\cal L}=\sqrt{-g} \Big(R + \fft{2}{\beta} (1 - \sqrt{1 + \beta F^2})\Big),
\ee
where $F=dA$ is the Maxwell field strength and $F^2=F^{\mu\nu} F_{\mu\nu}$. We ignore the $(\epsilon^{\mu\nu\rho\sigma} F_{\mu\nu} F_{\rho\sigma})^2$ term since it gives no contribution to equations of motion for purely electric ansatz. The limit of the coupling parameter $\beta\rightarrow 0$ yields EM gravity. Note that for convenience we take $\beta=1/b$ as our coupling constant, where $b$ is typically used in literature.

\subsection{Electrically charged black hole}

The theory admits an exact solution of electrically charged spherically symmetric and static black hole
\bea
ds^2 &=& - f(r) dt^2 + \fft{dr^2}{f(r)} + r^2 d\Omega_2^2\,,\qquad A=a(r) dt\,,\nn\\
f &=& 1 - \fft{2M}{r} + \fft{r^2-\sqrt{r^4 + 2\beta Q^2}}{3\beta}
+ \fft{4Q^2}{3r^2} {}_2F_1[\ft14,\ft12;\ft54,-2\beta Q^2/r^4]\,,\nn\\
a &=& -\frac{Q}{r} \, _2F_1[\ft{1}{4},\ft{1}{2};\ft{5}{4};-2\beta Q^2/r^4]\,.
\eea
The solution reduces to the standard RN solution in the $\beta\rightarrow 0$ limit.
The matter energy-momentum tensor $T^{\mu}_\nu={\rm diag}\{-\rho,p_r, p_T\}$ of the BI field is given by
\be
\rho = -p_r =  \fft{\sqrt{2 \beta Q^{2} + r^{4}} - r^{2}}{\beta r^{2}}\,,\qquad
\rho+p_T =  \fft{2 Q^{2}}{r^{2} \sqrt{2 \beta Q^{2} + r^{4}}}\,.
\ee
Therefore we have also
\be
\rho - P_{T} =\fft{(\sqrt{2 \beta Q^{2} + r^{4}} - r^{2})^{2}}{\beta r^{2} \sqrt{\beta Q^{2} +r^{4}}}\,,\qquad
\rho + p_{r} + 2p_{T} =  \fft{2(\sqrt{2 \beta Q^{2} + r^{4}} - r^{2})}{\beta \sqrt{2 \beta Q^{2} + r^{4}}}\,.
\ee
This implies that for $\beta\ge 0$, both strong and dominant energy conditions are satisfied. For $\beta <0$, the dominant energy condition is violated, but the strong and weak energy conditions remain satisfied.

The solution is asymptotic to Minkowski spacetime with the RN black hole as the leading falloffs
\be
f\sim 1 - \fft{2M}{r} + \fft{Q^2}{r^2} + \cdots\,,\qquad a\sim -\fft{Q}{r} + \cdots\,.
\ee
We have chosen the gauge choice for the Maxwell field that the electric potential vanishes at asymptotic infinity.

The solution has two integration constants $(M,Q)$, describing the mass and electric charge of the spacetime. For sufficiently large $M$, there is an event horizon $r_+$ which is the largest root of $f(r)$. The temperature, entropy and electric potential are
\be
T = \frac{r_+^2 +\beta -\sqrt{r_+^4 +2 \beta  Q^2}}{4 \pi  \beta  r_+}\,,\qquad
S=\pi r_+^2\,,\qquad \Phi=\frac{Q \, _2F_1\left(\frac{1}{4},\frac{1}{2};\frac{5}{4};-\frac{2 Q^2 \beta }{r_+^4}\right)}{r_+}\,.
\ee
It is easy to verify that the first law $dM=TdS + \Phi dQ$ is satisfied for any $f(r_+)=0$. For fixed $(M,Q)$, we find that
\be
S(M,Q,\beta) = S_0 + \beta S_1 + {\cal O}(\beta^2)\,,
\ee
with
\be
S_0 = \pi \left(\sqrt{M^2-Q^2}+M\right)^2,\qquad
S_1 = \fft{\pi}{10}Q^2\Big(\frac{\left(4 \mu ^2-3\right) \mu }{\sqrt{\mu ^2-1}}+1 -4 \mu ^2\Big)\,,\qquad \mu = \fft{M}{Q}\ge 1\,.
\ee
Thus the requirement that $\Delta S=\beta S_1\ge 0$ imposes that $\beta \ge 0$, in which case, the matter energy-momentum tensor satisfies both the strong and dominant energy conditions.

For $T\ge 0$, one must have
\be
r_+\ge \sqrt{Q^2- \ft12\beta}\,.\label{rpbe}
\ee
The saturation of the inequality gives rise to the extremal limit with $T=0$.
In this limit, mass is no longer an independent parameter, but a function of the charge $Q$:
\bea
M_{\rm ext} &=& \frac{2 Q^2-\beta+4 Q^2 \, _2F_1\left(\frac{1}{4},\frac{1}{2};\frac{5}{4};-\frac{8 Q^2 \beta }{\left(\beta -2 Q^2\right)^2}\right)}{3 \sqrt{4 Q^2-2 \beta }}\cr
&=&  Q-\frac{\beta }{20 Q}-\frac{\beta ^2}{288 Q^3}-\frac{\beta ^3}{1664 Q^5}+O\left(\beta ^4\right)\,.
\eea
Thus we see that at the linear $\beta$ order, $\Delta S>0$ requires $\beta>0$ and consequently $\Delta M<0$, i.e.~gravity becomes weaker than the electric force under the perturbation. Thus we see that at the linear order of $\beta>0$, the BI theory satisfies the criteria of \cite{Cheung:2018cwt} as a good effective theory.
For this reason, we focus on the $\beta>0$ case in this paper, but we also discuss what happens when $\beta$ is negative.

\subsection{Charged scalar perturbation}

A charged complex scalar under gauge field $A$ satisfies the charged Klein-Gordan equation
\be
(D^\mu D_\mu - m^2) \Phi =0\,,\qquad D_\mu =\nabla_\mu - {\rm i} q A_\mu\,,
\ee
where $(m,q)$ are the mass and charge of the fundamental scalar $\Phi$. It is a standard procedure to consider separation of variables, namely $\Phi= e^{-{\rm i} \omega t} R(r) Y_{\ell,m}(\theta,\varphi)$, where $Y_{\ell,m}$ are spherical harmonics. The radial function $R(r)$ satisfies
\be
-\Delta \fft{d}{dr} \Delta \fft{d}{dr} R + UR=0\,,\label{starteq}
\ee
with
\be
\Delta = r^2 f\,,\qquad U(r) = \Delta\, (m r^2 + \ell(\ell+1)) - r^4 (\omega + q a)^2\,.
\ee
We now study the boundary conditions both in the asymptotic infinity $(r\rightarrow \infty)$ and on the horizon $r=r_+$. The classical condition on the black hole horizon is that the scalar wave must be ingoing. It is instructive to introduce the tortoise coordinate $r^*$ by $dr^* = dr/f$.  For the extremal black hole we considered earlier, we have $r^*\rightarrow -\infty$ on the horizon, and
\be
r R \sim e^{\pm (\omega-\omega_c) r^*}\,,\qquad r^*\rightarrow -\infty\,,
\ee
where $\omega_c$ is
\be
\omega_c = -q a(r_+)= q \Phi\,.\label{omegacdef}
\ee
The ingoing condition selects the minus sign in ``$\pm$'' above.  In asymptotic region, for $m>\omega$, the scalar has the Yukawa falloff, namely
\be
R\sim \fft{e^{-\sqrt{m^2-\omega^2}\, r}}{r}\,,\qquad r^*\sim r\rightarrow \infty\,.
\ee
In other words, the wave function is bounded from the asymptotically infinity. Such a modes is called a quasi-bound states (QBS), and the frequency $\omega$ is necessarily discrete and complex \cite{Mai:2021yny}, i.e.
\be
\omega=\omega_r + {\rm i}\, \omega_i\,,\label{complexomega}
\ee
The real quantity $\omega_c$, depending on the parameters of the black hole and scalar equation, is a critical value such that superradiant effects take place for $\omega_r<\omega_c$. Thus the necessary condition for a QBS with superradiant effect is
\be
\omega < m\qquad\hbox{and}\qquad \omega_r <\omega_c\,.\label{condforqbs}
\ee
If the imaginary part of $\omega$, i.e.~$\omega_i$, were negative, then the QBS would be stable. However, it was shown \cite{Mai:2022thu} that superradiant QBS's, satisfying \eqref{condforqbs}, necessarily have positive $\omega_i$, and hence they are necessarily unstable. The focus of this paper is to search for the unstable QBS's for a wide range of the $\beta$ parameter.

\subsection{Necessarily conditions for unstable QBS's}

To examine the necessary condition for QBS, it is instructive to cast the scalar radial wave equation \eqref{starteq} into the Schr\"odinger form. To do this, we define $ \bar{R}=\Delta^{\fft12} R $, and \eqref{starteq} becomes
\be
-\fft{d^{2} \bar R }{d r^{2}}+V_{\rm eff} \bar R=\omega^{2}\bar R\,,
\ee
where the effective Schr\"odinger potential is
\bea
V_{\rm eff} &=& \omega^{2}-\fft{U}{\Delta^{2}} + \fft{2 \Delta'' \Delta -{\Delta'} ^2}{4 \Delta^{2}}\,,\nn\\
&=& \omega^{2} + \fft{m^{2}}{f} - \fft{(\omega + q a(r))^{2}}{f^2} + \fft{l (l+1)}{r^{2} f} + \fft{f^{''}}{2 f} - \fft{f^{'2}}{4 f^{2}} + \fft{f^{'}}{r f}\,.
\eea
The range of this effective potential is from horizon $r_+$ to asymptotic infinity.  In this paper, we focus on the extremal black hole, for which, we have
\be
	V_{\rm eff} \sim \left\{
	\begin{array}{ll}
		-\fft{\alpha_{+}}{(r-r_{+})^{4}}\,,&\qquad r \rightarrow r_{+}\,,\qquad \alpha_+ = \fft{4(\omega-\omega_c)^2}{f''(r_+)^2}\,;\nn\\
		m^{2}+\fft{\alpha_{\rm inf}}{r}\,,&\qquad r \rightarrow \infty\,,\qquad \alpha_{\rm inf}=M_{\rm ext} (m^2-2\omega^2) + 2 Qq\omega\,.
	\end{array}
	\right.
\ee
The existence of QBS requires that $V_{\rm eff}$ have a potential well. This however is not always possible under the superradiant QBS condition \eqref{condforqbs}. For example, for the RN black hole with $M_{\rm ext}=Q$ and $\omega_c=q$, it follows that $V_{\rm eff}$ has no potential well, but only a maximum. In particular, one has $\alpha_{\rm inf}>0$ under the condition \eqref{condforqbs}.

\begin{figure}[htp]
\centering
\includegraphics[width=350pt]{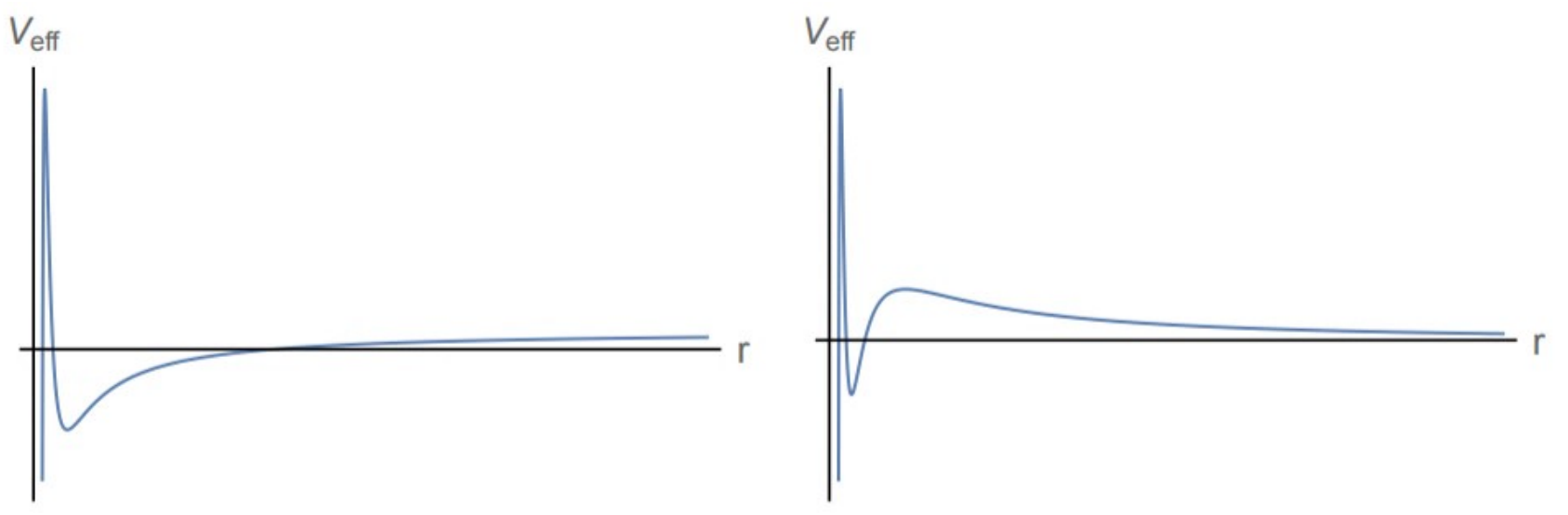}
\caption{\small There could be two types of potential wells. $\alpha_{inf} $ is negative in the first potential and positive in the second potential. We find that only the first type exists for EBI black hole.}\label{twopotential}
\end{figure}

In string theory, the RN black hole is a fine-tuned object of four-charge black holes in STU supergravity model. The additional charge parameters in the STU black hole implies that potential well could arise.  There are two types, as shown in Fig.
\ref{twopotential} \cite{Mai:2021yny}

The situation for EBI charged black hole is far more complicated, and we find that for general $\beta$, the potential well of the second type does not exist, but the first type does, which requires that $\alpha_{\rm inf}<0$. Note that $\alpha_{\rm inf}$ is an convex hyperbolic function of $\omega$, which has two roots
\be
\omega_\pm = \fft{q Q \pm \sqrt{q^2 Q^2 + 2m^2 M_{\rm ext}^2}}{2M_{\rm ext}}\,,\qquad \rightarrow\qquad \omega_-<0<\omega_+\,.
\ee
The condition $\alpha_{\rm inf}<0$ requires that $\omega>\omega_+$. However, the superradiant condition requires that $\omega<\omega_c$. Thus the existence of the first-type trapping well potential requires
\be
\omega_+ <\omega <\omega_c\,.\label{omegarange1}
\ee
We find that for suitable parameters, we indeed have $\omega_c/\omega_+>0$. For example, for $m=\omega_c$, we have
\be
\tilde\omega_+=\fft{\omega_+}{\omega_c}=\frac{3 \left(\sqrt{\frac{8}{9} ((\tilde \beta +2) H(\tilde\beta )+1)^2+\frac{1}{H(\tilde \beta )^2}}+\frac{1}{H(\tilde \beta )}\right)}{4 ((\tilde \beta +2) H(\tilde \beta )+1)}\,,\label{wpwc}
\ee
where
\be
H(\tilde\beta)=\, _2F_1\left(\frac{1}{4},\frac{1}{2};\frac{5}{4};-\tilde \beta  (\tilde \beta +2)\right)\,,
\ee
Note that we have introduced the dimensionless coupling constant $\tilde \beta$
\be
\tilde \beta = \fft{\beta}{r_+^2}\,.\label{dimensionlessbeta}
\ee
Under this parametrization and notation, we have
\be
M_{\rm ext}=\ft13 r_+ (1 + (\tilde \beta+2) H(\tilde\beta))\,,\qquad Q_{\rm ext}=
r_+ \sqrt{1 + \ft12 \tilde \beta}\,,\qquad
\omega_c =q\sqrt{1 + \ft12\tilde\beta}\, H(\tilde \beta)\,.\label{mqomegac}
\ee
Note from \eqref{rpbe} that for fixed $Q$, $\beta$ must be bounded above. However, in the above parametrization, we have fixed $r_+$, in which case, $\tilde \beta$ is unbounded above, but bounded below. It is easy to establish as $\tilde \beta$ runs from 0 to infinity, the ratio $\tilde \omega_+$ monotonously decreases from 1 to 0.958176.

Of course, to have simply $\alpha_{\rm inf}<0$ does not guarantee that $V_{\rm eff}$ yields a trapping well; it could be a monotonous function of $r$. This is indeed the case for some large $\omega>\omega_c$. We find however that for the $\omega$ lying in the range \eqref{omegarange1}, $V_{\rm eff}$ always gives a trapping well potential of the first type. Since the potential well of the first type can exist for small arbitrarily small $\beta$, we examine whether we can consider linear approximation of $\beta$ so that we can simplify the radial scalar equation. The superradiant QBS condition \eqref{condforqbs} can be imposed by parameterized
$(m,\omega)$ in terms of two positive dimensionless parameters $(x>0,y>0)$:
\be
\omega_c =  (1 + x)\omega\,,\qquad m^2 =(1 +y)\omega^2\,.
\ee
We now have
\be
\alpha_{\rm inf}=\fft23 r_+\omega^2 \left(
(\tilde \beta +2) (x-1) H(\tilde\beta)+\frac{3 (y+1)}{H(\tilde\beta)}+x-1\right).
\ee
For small $\tilde \beta$, we have
\be
\alpha_{\rm inf} = \fft23 r_+\omega^2\Big(3 (x+y)+\ft{3}{5} \tilde\beta  (x+y)+\ft{1}{75} \tilde\beta ^2 (-5 x-6 y-1)+O\left(\tilde \beta ^3\right)\Big).
\ee
Thus for fixed $(x,y)>0$, $\tilde \beta$ cannot be too small to achieve negative $\alpha_{\rm inf}$.  However, $\tilde \beta$ can be small when the $x$ and $y$ parameters behave like $(x,y)\sim (\tilde x, \tilde y) \beta^2$, in which case, we have
\be
\alpha_{\rm inf} = \fft23 r_+\omega_c^2 \tilde \beta ^2 \left(3 (\tilde x+ \tilde y)-\ft{1}{75}\right)+O\left(\tilde \beta ^3\right).
\ee
Thus we need small $(x,y)$ if we choose small $\tilde \beta$. In fact, if we set $x=y=0$, then we have
\be
\gamma = -\ft23 \omega_c^2 r_+\Big(1+ (\beta +2) H(\tilde \beta)-\frac{3}{H(\tilde\beta)}
\Big).
\ee
This quantity vanishes at $\beta=0$, but is always negative for $\tilde \beta >0$. Thus the trapping well potential always exists for arbitrarily small but non-vanishing $\beta$; however, it does not occur at the linearized level of $\tilde\beta$. The nonlinear effect of BI action plays an essential rule. Our numerical results of small $\tilde \beta$ will confirm this indeed.

The above analysis was based on the assumption that $\omega$ is real. As we mentioned earlier, it was proven that QBS must have complex frequency \eqref{complexomega}. In practice, as we shall see from our numerical results, $\omega_i \ll \omega_r$; therefore, the imaginary part of the $\omega$ does not affect the essence of the above discussion.

\section{Numerical methods}
\label{sec:methods}

In the previous section, we obtained the necessary condition for the unstable QBS's. However, having a trapping-well type of potential does not necessarily imply the existence of a bound state. We need actually find this state. In this paper, we adopt two rather different numerical methods to solve the radial charged scalar wave equation \eqref{starteq} for unstable QBS's. One is the shooting method and the other is the Chebyshev spectral method. The reason to adopt two methods is that the imaginary part of the QBS are infinitesimally small and we would like to use two independent methods to crosscheck our results.

\subsection{Shooting method}

Although the equation \eqref{starteq} cannot be solved exactly, it can be expressed in terms of power series to arbitrarily higher orders around the horizon $r_+$ or at the asymptotic infinity. For the QBS boundary conditions, the function $R(r)$ at the two boundaries behave like
\bea
	R \sim
	\left\{
	\begin{array}{ll}
		(r-r_{+})^{-2 i \chi_{2}}  e^{i \fft{(\omega-\omega_{c})r_{+}^{2}\chi_{1}}{r-r_{+}}}   \sum_{i=0}^{+\infty}a_{i}(r-r_{+})^{i} \,,\qquad
		r \rightarrow r_{+}\,,\nn\\
		r^{- \chi_{3}}   e^{-\sqrt{m^{2}-\omega^{2}} r}   \sum_{i=0}^{+\infty}\fft{b_{i}}{r^{i}}\,,\qquad
		r \rightarrow +\infty\,.
	\end{array}
	\right.
\eea
where
\bea
	\chi_{1} &=& \fft{ 2 }{r_{+}^{2}} \fft{ 1 }{  f^{''}(r_{}) } = \fft{r_{+}^{2} + \beta}{r_{+}^{2}} \,,\qquad
	\chi_{2} = \fft{q \sqrt{2 r_{+}^{2}+\beta}}{2 \sqrt{2}} + \fft{3 r_{+}^{4} + 3 r_{+}^{2} \beta + \beta^{2} }{3 r_{+} (r_{+}^{2} + \beta)}(\omega-\omega_{c})\,,\nn\\
	\chi_{3} &=& 1 - 2 i \chi_{2} + \fft{  2r_{+}(m^{2} - 2 \omega^{2}) + 3\sqrt{2} q \omega \sqrt{2 r_{+}^{2} + \beta} }{6 \sqrt{m^{2}-\omega^{2}}}\nn\\
	& &+\fft{ (m^{2} - 2 \omega^{2})( 2 r_{+}^{2} + \beta ) _{2}F_{1} [\ft12,\ft14;\ft54,-\fft{\beta (2 r_{+}^2 + \beta) }{r_{\rm h}^{4}}]}{3 r_{\rm h} \sqrt{m^{2}-\omega^{2}}}\,.\label{chi123}
	\eea
The coefficients $(a_n,b_n)$ of the power series are determined by solving the equation order by order, in terms of the leading coefficients $(a_0,b_0)$, both of which can be set to 1 for our linear differential equations.

We can use these two power series expansions at the both ends as spatial ``initial'' condition and integrate outwards from horizon or inwards from some large $r$ to obtain functions $(R_+, R_-)$ respectively. We then require that these two functions match in the intermediate middle region $r=r_i$. As a linear function, $R_+(r_i)=R_-(r_i)$ can always be arranged by adjusting $(a_0,b_0)$ coefficients. The matching of their derivative gives the nontrivial condition. The Wronskian condition that is independent of $(a_0,b_0)$ is given by
\be
W(\omega)=\fft{R_+'}{R_+} - \fft{R_-'}{R_-} =0\,.
\ee
This condition leads to a discrete set of complex $\omega$ of QBS's if they exist. The implementation of the shooting method numerically is rather straightforward. We find that the real part of the unstable QBS's are fairly straightforward to obtain, but the imaginary part can be only determined with sufficient accuracy for intermediate values of $\beta$. For those values we can determine in high accuracy, we can perform crosscheck with those determined by the spectral method which we discuss below.

\subsection{Chebyshev spectral method}

The use of Chebyshev spectral method to find QBSs or QNMs was well documented in literature \cite{Grandclement:2007sb,Pani:2013pma}.  In order to apply for the extremal charged black hole of this paper, we change the radial coordinate from $r$ to $ x=\fft{r-r_{+}}{r} $, and the dimensionless coordinate $ x \in [0,1] $. The boundary conditions are
	\bea
	R \sim
	\left\{
	\begin{array}{ll}
		x^{-2 i \bar{\chi}_{2}} e^{i \fft{(\omega - \omega_{c}) (1-x)r_{+}\bar{\chi}_{1}}{x}}\,,\qquad
		x \rightarrow 0\,,\nn\\
		(1-x)^{\bar{\chi}_{3}} e^{-\sqrt{m^{2}-\omega^{2}} \fft{r_{+} x}{1-x}}\,,\qquad
		x \rightarrow 1\,,
	\end{array}
    \right.\nn\\
    \bar{\chi}_{1}=\chi_{1}\,,\qquad
    \bar{\chi}_{2}=\chi_{2}\,,\qquad
    \bar{\chi}_{3}=\chi_{3} + 2 i \chi_{2}\,.
    \eea
Where $ \chi_{1} , \chi_{2}, \chi_{3}$ are from (\ref{chi123}). The boundary conditions show that the function $R$ oscillates infinitely at both $ x=0 $  and it exponentially decays at $x=1$. We can take out these extreme behaviors by redefining the radial function
\be
R= x^{-2 i \bar{\chi}_{2}} e^{i \fft{(\omega - \omega_{c}) (1-x)r_{+}\bar{\chi}_{1}}{x}} (1-x)^{\bar{\chi}_{3}} e^{-\sqrt{m^{2}-\omega^{2}} \fft{r_{+} x}{1-x}}\, u(x)\,.
\ee
Here we also have multiplied the equation by $ \fft{(x-1)^{2}}{x^{2}} $ to ensure that the coefficients of the each derivatives do not diverge or become zero at both $x=0$ and $x=1$. Then we obtain the equation for the function $u$ in the $x$ coordinate.
\be
c_{2}(\omega,x) u''(x) + c_{1}(\omega,x) u'(x) + c_{0}(\omega,x) u(x) = 0\,.\label{uequ}
\ee
The goal is to solve this better behaved equation. In spectral methods, one can approximate a function by a finite sum of certain appropriately ($\omega$)-measured orthonormal polynomials
\be
u(x)=\sum_{i=1}^{N} \bar{u}_{i} P_{i}(x)\,,\qquad \int dx\, \omega(x) P_i(x) P_j(x)=\delta_{ij}\,.
\ee
This can be shown to be equivalent to constructing an interpolation polynomial in a grid with $N$ points, namely
\be
u(x) = \sum_{i=1}^{N} u(x_{i}) l_{i}(x)\,,\qquad l_{i}(x) = \prod_{j=0,j \neq i}^{N} \left(\fft{x-x_{j}}{x_{i} - x_{j}}\right).
\ee
An equidistant grid may be the simplest choice, but it suffers from the Runge problem. A better approach is to conduct the interpolation within a non-uniform grid, and the optimal choice is to use Chebyshev points $x_i$ defined by the extremum points of Chebyshev polynomials $T_i$, namely
	\be
	x_{i} = \fft12 + \fft12 \cos(\fft{(i-1) \pi}{N})\,,\qquad
	T_{i}(x) = \cos ((i-1) \cos^{-1} x)\,, \qquad
	i=1,2,3...\,.
	\ee
	The interpolation maps $ u(x) \rightarrow u=\left(u(x_1),u(x_2),....,u(x_N)\right) $, and the equation \eqref{uequ} is transformed into an algebraic system for the $N$-dimensional vectors, as the differential operaters $ \fft{d}{d x},\fft{d^{2}}{d x^{2}} $ transform to the differential matrixs $ D_{1}^{N},D_{2}^{N} $. As long as the boundary conditions at $u(x_1) $ and $u(x_N)$ are taken into account, the problem reduces to an algebraic eigenvalue problem in $N \times N$ dimensions, which can be solved by standard methods \cite{Grandclement:2007sb,Pani:2013pma}, such as Newton iteration method.

Our goal is to find the discrete set of unstable QBS's satisfying the appropriate boundary conditions for give $\beta$. To perform the numerical calculation using the spectral method, we need specify the grid point number $N$. If the QBS's exist, it will yield a complex frequency that depends on $N$, i.e.~$\omega(N)$. When $N$ increases, the value of $\omega(N)$ typically converges quickly. Therefore, we obtain the numerical result with desired accuracy for a sufficiently large $N$.
For many such calculations in literature, black hole metrics are given in terms of polynomials or rational polynomials of the coordinates, where it is commonly sufficient to set $N$ to be a few hundreds. In our case, the black hole metric involves a hypergeometric function and it is therefore inherently more complicated. We find that our results requires larger $N$. The convergent of the real part is relatively fast and it requires typically $N$ be over 1000, but not necessary be over 2000. The convergent of the imaginary part can be significantly slower. For large or intermediate values of $\beta$, a few thousands would be sufficient for $N$. Our computer power sets a limit of $N=10000$. However, for small $\beta$, the convergence requires $N$ be hundreds of thousands. In this case, we can try to use data extrapolation to yield a rough estimate. However, the explicit accurate result of the imaginary frequency is not so essential since we can apply the powerful theorem established in \cite{Mai:2021yny} that states, if the real part of frequency satisfy the superradiant condition, the QBS is necessarily unstable with a positive imaginary part of frequency.

\section{Numerical results}
\label{sec:results}

Numerical calculations require us to set parameters of the equations to some specific values. The extremal charged black hole is specified by two parameters, the horizon radius and BI coupling parameter $\beta$. After introducing the dimensionless $\tilde\beta$ in \eqref{dimensionlessbeta}, we can set $r_+=1$ without loss of generality. In this case, as was discussed earlier around \eqref{mqomegac}, the parameter $\tilde \beta$ is bounded below, but unbounded above.  The linear scalar wave equation involves three parameters, the dimensionless $\ell$ and two dimensionful $(\omega_c, m)$. For simplicity, we consider only the case with
\be
\omega_c=m\,,\qquad \ell=1\,.
\ee
It is thus useful to introduce the dimensionless complex frequency, defined by
\be
\tilde\omega=\tilde \omega_{\rm r} + {\rm i} \tilde \omega_{\rm i} \equiv \fft{\omega}{\omega_c}=\fft{\omega_{\rm r}}{\omega_c} + {\rm i}\fft{\omega_{\rm i}}{\omega_c}\,.\label{omeganodim}
\ee
Our task is to search for the unstable QBS's $(\omega_{\rm i}>0)$ numerically with one free dimensionless parameter $\tilde \beta$. It was established that such a superradiant QBS does not exist when $\tilde \beta=0$, corresponding to the extremal RN black hole. We focus on the case with $\tilde \beta>0$. Our data indicate a general conclusion that unstable QBS's exist for all $\tilde\beta$. Since the real part of $\tilde \omega$ of  the unstable QBS's can be very close to 1, it is useful to define $\eta$, by
\be
\eta=1 - \tilde \omega_{\rm r}\,.\label{eta}
\ee

\subsection{Intermediate $\tilde \beta$}

We shall first present the detailed results of the unstable QBS's with two specific coupling $\tilde\beta$ of the intermediate size, namely $\tilde \beta=10$ and $\tilde \beta=20$. In each case, there exist more than just one unstable QBS, and these discretized unstable QBS's can be characterized by the overtone number $n$, which counts the number of the peaks of the radial wave function $|R(r)|$. The $n=1$ solution can be called the ground state while the higher overtone solutions are excited states. In Table 1, we give the dimensionless complex frequency $\tilde \omega$ for some unstable QBS's with low-lying overtone number. Note that our calculation allows us to obtain accuracy over more than 10 significant figures for some of these states, but we shall present data only with five or six significant figures in this paper, so as not to overflow the text.

\begin{center}
\begin{tabular}{|c|c|c|}
  \hline
  $n$ & $\tilde\beta=10$ & $\tilde\beta=20$ \\ \hline
  1 &  $0.99947 + 1.94607\times 10^{-43} {\rm i}$ & $0.99824 + 1.10193\times 10^{-47} {\rm i}$ \\
 2& $0.99972+ 5.18693\times 10^{-47} {\rm i}$ & $0.99905 + 2.10225\times 10^{-52}{\rm i}$ \\
 3 & $0.99984 + 9.49056\times 10^{-50}{\rm i}$ & $0.99941 + 4.24260 \times 10^{-55}{\rm i}$ \\
4 & $0.99989 + 6.03056\times 10^{-52} {\rm i}$ & $0.99960 + 7.56974\times 10^{-58}{\rm i}$\\
5 & $0.99992 + 8.94163 \times 10^{-54} {\rm i}$   & $0.99971 + 3.54886\times 10^{-60}{\rm i}$ \\
6 &$0.99994+ 2.43761\times 10^{-55}${\rm i}  & $0.99978 +  3.43055\times 10^{-62}{\rm i}$\\
  \hline
\end{tabular}
\bigskip

\small{Table 1. The dimensionless complex frequencies $\tilde\omega$ of \eqref{omeganodim} for unstable QBS's with low-lying overtones.}
\end{center}

We see that the imaginary part $\tilde \omega_{\rm i}$ in Table 1 are exceedingly small, which is a general feature of all the unstable QBS's. This is the primary reason why we adopt both shooting and spectral methods for our numerical calculations, so that we can perform crosscheck. Generally speaking, for low-lying overtone numbers, both methods are equally capable to obtain the real part $\tilde \omega_r$ in higher accuracy. The spectral method is better at getting the accurate imaginary part, for intermediate or large $\tilde\beta$. In Table 1, the $\tilde \omega_{\rm i}$ values are largely obtained by the spectral method, and verified with appropriate accuracies by the shooting method. In particular, the ground state $(n=1)$ with $\tilde\beta=10$ was double checked by both the shooting and spectral methods, with accuracy up to 10 significant figures for both real and imaginary parts.

\begin{figure}[htp]
\centering
\includegraphics[width=200pt]{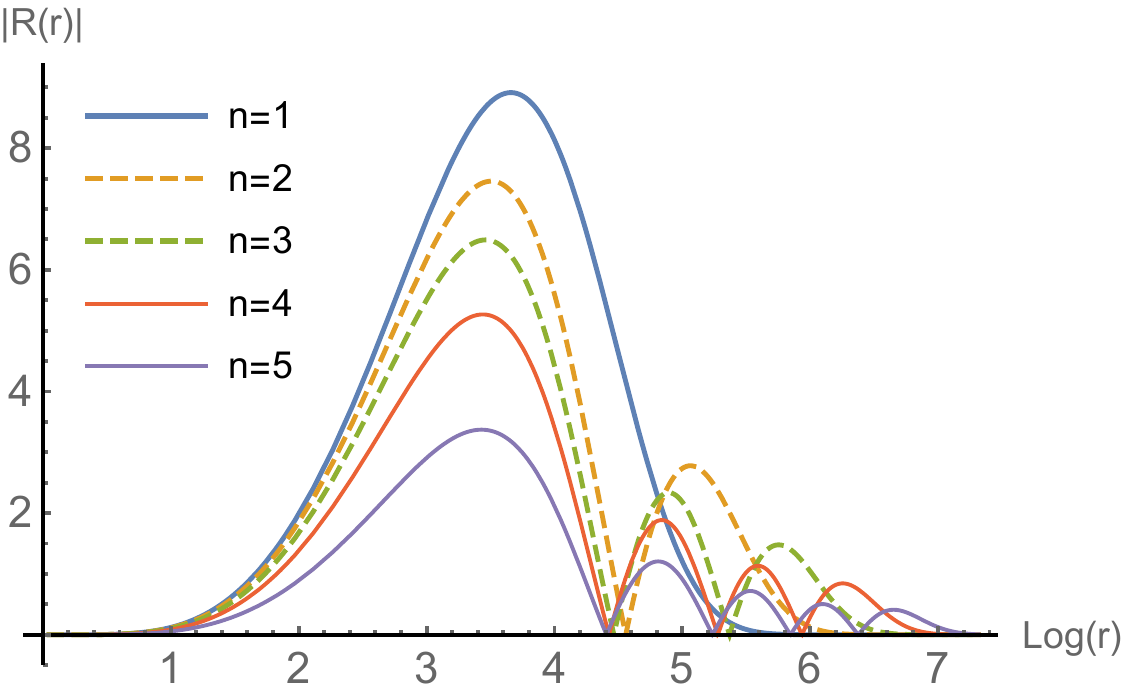}\ \
\includegraphics[width=200pt]{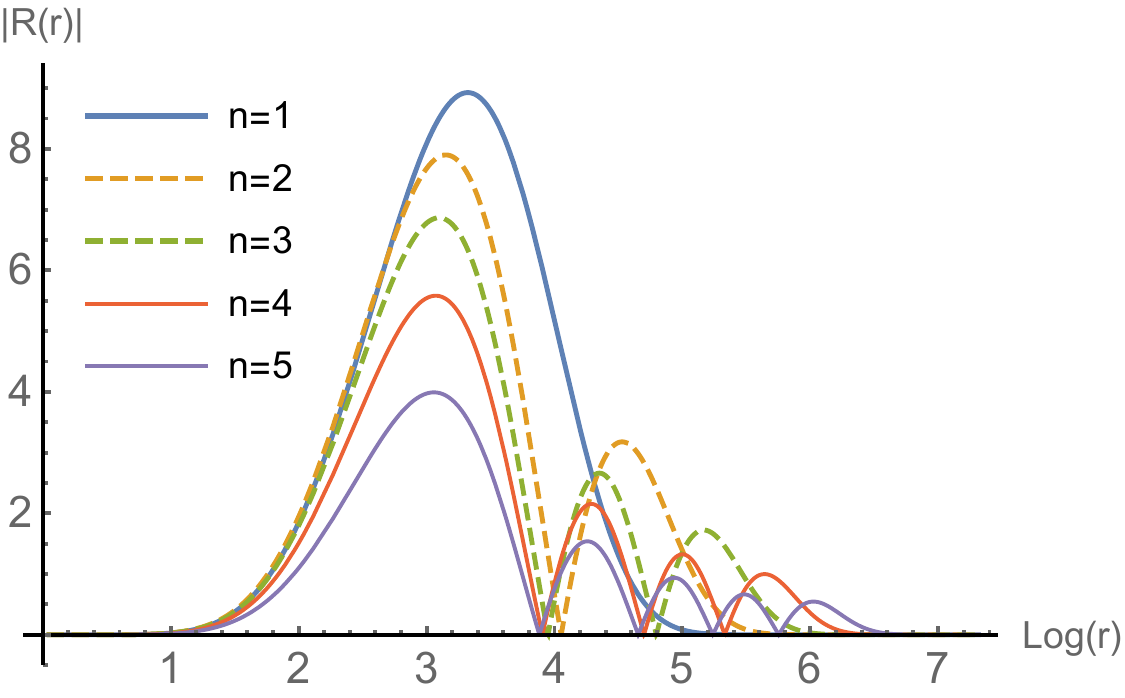}
\caption{\small Examples of unstable QBS's with low-lying overtone number $n$. To fit all the radial wave functions in one plot for each $\tilde \beta$, we use logarithmic scale of the radial coordinate. The (linear) functions are scaled so that they give a better viewing. The left panel has $\tilde\beta=10$ and the right has $\tilde\beta=20$. }\label{radialRplot}
\end{figure}

A further check on the unstable QBS's listed in Table 1 is to examine the shape of the radial wave function $R(r)$. For overtone $n$, the quantity $|R(r)|$ should have $n$ peaks. We draw the $|R(r)|$ for low-lying $n$  examples in Fig.~\ref{radialRplot}. In order to fit graphs in one plot for each $\tilde\beta$, we use the logarithmic scale of the $r$ coordinate. Note that the (linear) function $R(r)$ for each $n$ is not normalized, but scaled appropriately for a better viewing effect.

An interesting question emerges. For given $\tilde \beta$, is the overtone number for excited unstable QBS's finite, or can there be an infinite number of such unstable states? One thing has already been established that for any unstable QBS, we must have $\tilde \omega^{(n)}_r<1$, for any given $n$. To be precise, it follows from \eqref{wpwc}
that we must have
\be
\tilde\beta=10:\quad
0.98844 <\tilde \omega^{(n)}<1\,;\qquad\tilde\beta=20:\quad
0.98287<\tilde \omega^{(n)}<1\,.
\ee
This implies that $|\omega_r^{(n+1)}-\omega_r^{(n)}|$ approaches zero as $n$ approaches infinity, even if such states with infinitely large overtone numbers actually exist. Therefore, it quickly becomes an impossible task to construct such states with large overtone number, since numerically they cannot be easily distinguished. In Table 2, we list only the real part of $\tilde \omega$ up to and including $n=12$ for both the $\tilde \beta=10$ and 20 cases. The data for $n\ge 6$ were obtained by the shooting method, since the spectral method becomes ineffective to distinguish two adjacent QBS's of large $n$.

\begin{center}
\begin{tabular}{|c|c|c|c|c|c|c|}
  \hline
  $n$ & 1 & 2 & 3 & 4 & 5 & 6 \\ \hline
  $\tilde\beta=10$ & 0.999437 & 0.999721 & 0.999835 & 0.999892 & 0.999924 & 0.999943\\
  $\tilde\beta=20$ & 0.998246 & 0.999049 & 0.999413 & 0.999604 & 0.999716 & 0.999786\\
  \hline
  $n$ & 7 & 8 & 9 & 10 & 11 & 12 \\ \hline
  $\tilde\beta=10$ & 0.999956 & 0.999965 & 0.999972 & 0.999978 & 0.999982 & 0.999986\\
  $\tilde\beta=20$ & 0.999834 & 0.999867 & 0.999891 & 0.999909 & 0.999927 & 0.999942\\
  \hline
  \end{tabular}
\bigskip

{\small Table 2. The real part $\tilde \omega_r$ of the complex frequency $\tilde \omega$ for the unstable QBS's of low-lying overtones.}
\end{center}

The data suggests that we can perform data-fitting on the quantity $\eta$ \eqref{eta} in terms of an inverse polynomial of $n$. For the best results, we find
\bea
\tilde\beta=10:&& 10^{4}\,\eta \sim -0.221856 +\frac{3.76694}{n}+\frac{7.00194}{n^2}-\frac{4.91735}{n^3}\,,\nn\\
\tilde\beta=20:&& 10^3\,\eta \sim -0.0862095 +\frac{1.5763}{n}+\frac{1.7555}{n^2} -\frac{1.49175}{n^3}\,.
\eea
Both formulae can produce less than 1\% of error for the 12 QBS's given in Table 2. What is important is that the leading terms are negative, implying that $n$ must be finite for $\eta>0$. For the above two examples, the maximum values of $n$ are 18 and 19 respectively. It is worth pointing out that if we consider polynomials up to a different order, we shall get less accuracy in data fitting, but the leading order was always negative. This of course is not a proof that the maximum $n$ is finite, but it is highly suggestive.

For now on, we shall focus only on the ground states ($n=1$), since the purpose of this paper is not to classify the QBS's, but to establish whether there exists an unstable QBS's for given $\tilde \beta$.  In Table 3, we list a variety of the ground unstable QBS's for intermediate values of $\beta$. All the real parts are verified by both the spectral and shooting methods. The imaginary parts were verified by both methods for the listed states with $\tilde \beta$'s lying from 1 to 10. The $\tilde \omega_{\rm i}$ values with $\tilde \beta$ less than 1 in Table 3 were obtained using the shooting method only, since the spectral method becomes very ineffective in computing the imaginary part ($\tilde \omega_{\rm i}$) for small $\tilde \beta$.

\begin{center}
\begin{tabular}{|c|c|c|c|c|c|}
  \hline
  $\tilde\beta$ & $\eta=1-\tilde\omega_{\rm r}$ & $\tilde\omega_{\rm i}$ &$\tilde\beta$ & $\eta=1-\tilde\omega_{\rm r}$ & $\tilde\omega_{\rm i}$ \\ \hline
  10 & $5.63046\times 10^{-4}$ & $1.94607\times 10^{-43}$ &
  9 & $4.59571\times 10^{-4}$ & $5.47987\times 10^{-43} $ \\ \hline
  8 & $3.62781\times 10^{-4}$ & $1.54294\times 10^{-42} $ &
  7 & $2.74135\times 10^{-4}$ & $4.29799\times 10^{-42} $ \\ \hline
  6 & $1.95261\times 10^{-4}$ & $1.16133\times 10^{-41} $ &
  5 & $1.27907\times 10^{-4}$ & $2.92884\times 10^{-41} $ \\ \hline
  4 & $7.38365\times 10^{-5}$ & $6.36014\times 10^{-41} $ &
  3 & $3.45716\times 10^{-5}$ & $9.84205\times 10^{-41} $ \\ \hline
  2.5&$2.07706\times 10^{-5}$ & $9.40010\times 10^{-41} $ &
  2  &$1.08156\times 10^{-5}$ & $6.34651\times 10^{-41} $ \\ \hline
  1.5&$4.46048\times 10^{-6}$ & $2.22304\times 10^{-41} $ &
  1.25&$2.48385\times 10^{-6}$ & $8.42189\times 10^{-42} $ \\ \hline
  1 &  $1.18535\times 10^{-6}$ & $1.85839\times 10^{-42} $ &
  0.9 &$8.29039\times 10^{-7}$ & $8.07937\times 10^{-43} $ \\ \hline
  0.8 &$5.52692\times 10^{-7}$ & $2.91427\times 10^{-43} $ &
  0.7 &$3.46609\times 10^{-7}$ & $8.21418\times 10^{-44} $ \\ \hline
  0.6 &$2.00556\times 10^{-7}$ & $1.65277\times 10^{-44} $ &
  0.5 &$1.03899\times 10^{-7}$ & $2.05663\times 10^{-45} $ \\ \hline
  0.4 &$4.58199\times 10^{-8}$ & $1.23607\times 10^{-46} $ &
  0.3 &$1.56472\times 10^{-8}$ & $2.22977\times 10^{-48} $ \\
  \hline
\end{tabular}
\end{center}

{\small Table 3. A list of ground states $(n=1)$ of the unstable QBS's for intermediate values of $\tilde \beta$. We see that the maximum value of $\tilde \omega_{\rm i}$ is in the vicinity of $\tilde \beta=3$.}

As we can see from Table 3, the real part $\tilde \omega_r$ or $\eta$ are monotonous function of $\tilde\beta$, but $\tilde \omega_{\rm i}$ is a convex function with a maximum occurring in the vicinity of $\tilde \beta=3$. We select three points $\tilde\beta =2.5,3,4$, and use a quadratic function of $\tilde \log\tilde \beta$ to data-fit $\log(\tilde\omega_{\rm i})$, we find
\be
\log\tilde \omega_{\rm i} \sim -92.11 - 3.765 (\log\tilde\beta - 1.041)^2 + {\cal O}((\log(\tilde\beta -1.041)^3)\,.\label{quadfitting}
\ee
According to this data-fitting function, the maximum value would occur at $\tilde \beta=1.041$, i.e.~$\tilde \beta=2.832$. We perform numerical calculation, and find the maximum imaginary part occurs at $\beta=2.833$, for which we have
\be
\tilde \omega_{\rm i}=(1 - 2.95275\times 10^{-5}) + 9.96443 \times 10^{-41}\,{\rm i}\,,\qquad \log (9.96443 \times 10^{-41})=-92.1107\,.
\ee
Thus we see that the data-fitting function using only three data actually produces a quite accurate value of $\tilde \beta$ for the maximum $\tilde \omega_{\rm i}$.  In the left panel of Fig.~\ref{alldata}, all the $\tilde \omega_{i}$'s listed in Table 3, together with further data given presently, are plotted as a function of $\tilde \beta$. The solid line describes the quadratic relation \eqref{quadfitting}. We see that it fits with quite a few further data away from the maximum point.

\subsection{Large $\tilde \beta$}

For large $\tilde \beta$, (e.g.~$\tilde \beta >10$), the spectral flow method turns out to be very efficient for both real and imaginary parts of the frequency. A grid point number $N\sim 1000,2000$ is quite sufficient in these cases. We list the complex frequencies in Table 4 for a large number of $\tilde \beta$, up to and including $\tilde \beta=200$. We plot these results, together with other data, in Fig.~\ref{alldata} to give a direct picture of these numerical results.
\bigskip

\begin{tabular}{|c|c|c|c|c|c|}
  \hline
  $\tilde\beta$ & $\eta=1-\tilde\omega_{\rm r}$ & $\tilde\omega_{\rm i}$
  &$\tilde\beta$ & $\eta=1-\tilde\omega_{\rm r}$ & $\tilde\omega_{\rm i}$\\ \hline
  200  & $1.04935\times 10^{-2}$ & $1.38327\times 10^{-88} $ &
  190  & $1.02976\times 10^{-2}$ & $5.19283\times 10^{-87} $ \\ \hline
  180  & $1.00887\times 10^{-2}$ & $2.14754\times 10^{-85} $ &
  170  & $9.86532\times 10^{-3}$ & $9.86013\times 10^{-84} $ \\ \hline
  160  & $9.62561\times 10^{-3}$ & $5.07060\times 10^{-82} $ &
  150  & $9.36748\times 10^{-3}$ & $2.95026\times 10^{-80} $ \\ \hline
  140  & $9.08841\times 10^{-3}$ & $1.96497\times 10^{-78} $ &
  130  & $8.78536\times 10^{-3}$ & $1.51860\times 10^{-76} $ \\ \hline
  120  & $8.45466\times 10^{-3}$ & $1.38379\times 10^{-74} $ &
  110  & $8.09179\times 10^{-3}$ & $1.51520\times 10^{-72} $ \\ \hline
  100  & $7.69116\times 10^{-3}$ & $2.03969\times 10^{-70} $ &
  90   & $7.24573\times 10^{-3}$ & $3.47083\times 10^{-68} $ \\ \hline
  80   & $6.74660\times 10^{-3}$ & $7.72761\times 10^{-66} $ &
  70   & $6.18230\times 10^{-3}$ & $2.35115\times 10^{-63} $ \\ \hline
  60   & $5.53805\times 10^{-3}$ & $1.03386\times 10^{-60} $ &
  50   & $4.79483\times 10^{-3}$ & $7.07490\times 10^{-58} $ \\ \hline
  40   & $3.92916\times 10^{-3}$ & $8.33147\times 10^{-55} $ &
  30   & $2.91666\times 10^{-3}$ & $1.94425\times 10^{-51} $ \\ \hline
  25   & $2.35173\times 10^{-3}$ & $1.28945\times 10^{-49} $ &
  20   & $1.75356\times 10^{-3}$ & $1.10193\times 10^{-47} $ \\ \hline
  15   & $1.14120\times 10^{-3}$ & $1.25824\times 10^{-45} $ &
  12.5 & $8.42938\times 10^{-4}$ & $1.51068\times 10^{-44} $ \\ \hline
\end{tabular}
\bigskip

{\small Table 4. For large $\tilde \beta$ values, the spectra method is very effective for finding the unstable QBS's of $n=1$, and grid point number $N=1000$ is sufficient. Here we list the complex frequencies $\tilde \omega$ for a variety of $\tilde \beta$.}

\subsection{Small $\tilde \beta$}

\begin{figure}[htp]
\centering
\includegraphics[width=190pt]{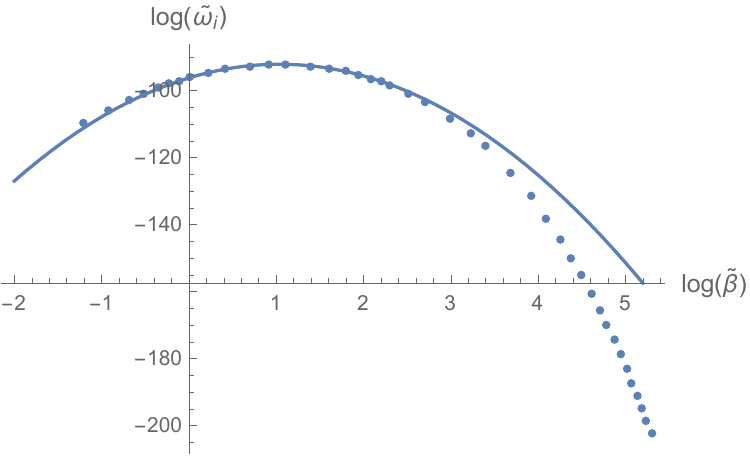}\ \ \
\includegraphics[width=190pt]{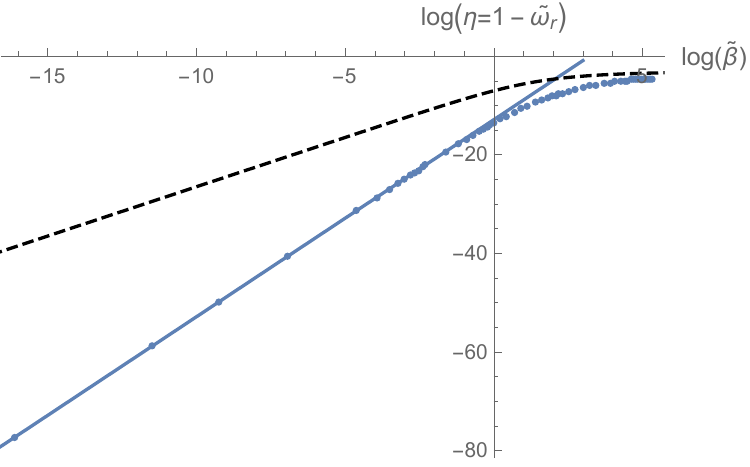}
\caption{\small The two panels show the all of our numerical data, pictured as dots, of the unstable ground ($n=1)$ QBS's. The left panel is about the real part $\tilde\omega_r$. The dashed line is the upper bound so that the exicited states lie between the dashed line and the line connecting the dots.}\label{alldata}
\end{figure}

For small $\tilde \beta$, both the spectral and shooting methods are ineffective in computing the imaginary part of the frequency. However, the real part can still be obtained with high accuracy, by both methods. In Table 5, we give only the real parts, via $\eta$, for a large span of the parameter $\tilde\beta$. Together with the earlier data presented, we have the real part for $\tilde \beta$ ranging from $10^{-7}$ to 200. These data are pictured as dots in the right panel of Fig.~\ref{alldata}.

\begin{center}
\begin{tabular}{|c|c|c|c|c|c|}\hline
	$\tilde \beta$ & 0.2 & 0.1 & 0.09 & 0.08 & 0.07 \\ \hline
	$\eta$ & $3.34458\! \times\! 10^{-10}$ & $2.26840 \!\times\! 10^{-10}$ & $1.50076 \!\times\! 10^{-10}$ & $9.44789 \!\times\! 10^{-11}$ & $5.58489 \!\times\! 10^{-11}$ \\
	\hline

$\beta$ & 0.06 & 0.05 & 0.04 & 0.03 & 0.02 \\ \hline
	$\tilde \eta$ & $3.04011\! \times\! 10^{-11}$ & $1.47856 \!\times\! 10^{-11}$ & $6.10785 \!\times\! 10^{-12}$ & $1.94911 \!\times\! 10^{-12}$ & $3.88319 \!\times\! 10^{-13}$ \\
	\hline
	
$\tilde \beta$ & $10^{-2}$ & $10^{-3}$ & $10^{-4}$  & $10^{-5}$ & $10^{-7}$ \\ \hline
	$\eta$ & $2.44793\! \times\! 10^{-14}$ & $2.46700 \!\times\! 10^{-18}$ & $2.46892 \!\times\! 10^{-22}$ & $2.46911 \!\times\! 10^{-26}$ & $2.46914 \!\times\! 10^{-34}$  \\
	\hline
\end{tabular}
\end{center}

{\small Table 5: The real part $\tilde \omega_r$ of the frequency of the unstable QBS's for small $\tilde\beta$, given as $\eta=1-\tilde\omega_r$. We see that $\eta$ becomes very small for $\tilde \beta=10^{-7}$, the last one in the list.}

As we can see in the left panel of Fig.~\ref{alldata}, for small $\tilde \beta\ll 1$, the numerical data, depicted as dots, behaves linearly. We select two data, associated with $\tilde \beta =10^{-7}$ and $\tilde \beta=10^{-5}$, and derive a linear relation between $\log(\eta)$ and $\log(\tilde \beta)$. We find that it is given by
\be
\log(\eta)= -12.9117 + 4.00000\log(\tilde\beta) + \cdots\,,\qquad \hbox{when}\qquad \tilde\beta \rightarrow 0\,.
\ee
This linear relation is drawn as the solid line in the right panel of Fig.~\ref{alldata}. We see that linear relation fits with the numerical data very well for $\tilde\beta <0.01$. This accurate numerical data fitting indicates that the coefficient of $\log(\tilde\beta)$ is 4 up to six significant figures. We can therefore confidently conjecture that it is exactly 4. This implies that for small $\tilde\beta$, the real part of the frequency is given by
\be
\eta = 2.46906\times 10^{-6}\, \tilde \beta^4 + \cdots\,,\qquad \tilde\beta \rightarrow 0\,.\label{smallbetaeta}
\ee

The dashed line in the right panel of Fig.~\ref{alldata} describes $\log(\eta_+)$, with $\eta_+=1 - \tilde\omega_+$, where $\omega_+$ is given by \eqref{wpwc}. In other words, the dashed line represents the boundary values of $\tilde \omega_r$ for unstable QBS's. While the dots are associated with ground states, the excited unstable QBS's with higher overtone $n$ should lie in the region between the dashed line and the line connecting the dots.

\subsection{Extremely small $\tilde\beta$}

\begin{figure}[htp]
\centering
\includegraphics[width=200pt]{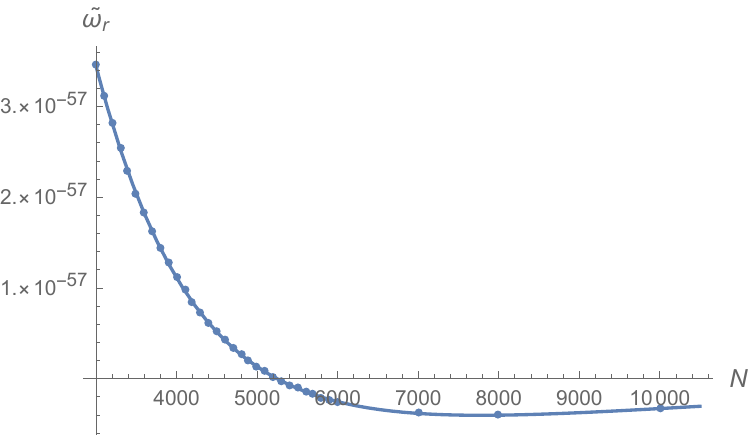}\ \
\includegraphics[width=200pt]{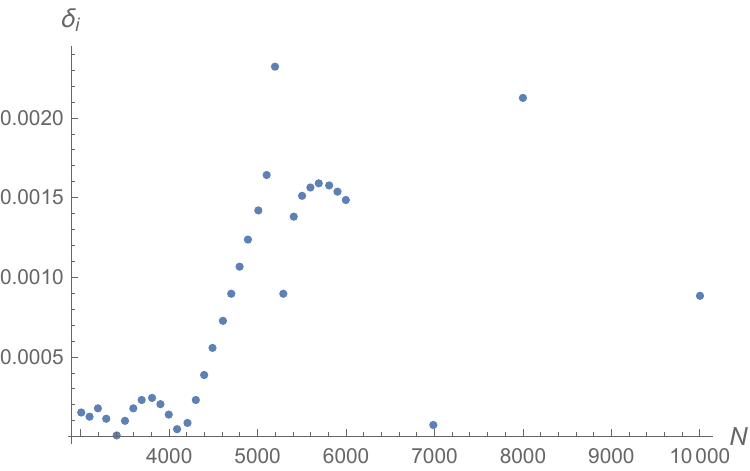}
\caption{\small }\label{extremesmall}
\end{figure}

As we have mentioned earlier, for extremely small $\tilde\beta$, i.e.~$\tilde\beta \le 10^{-5}$, the real part of the complex frequency $\tilde \omega$ can be still easily determined as \eqref{smallbetaeta}. The imaginary term is much tougher. Here we discuss in some detail the calculation for $\tilde \beta=10^{-7}$ case, where we obtained the results for various grid point number $N$, from 1000 to 10000, which becomes the stretch of our computer power. The real part converges quickly to the one given in the previous subsection. The imaginary part refuses to converge yet, as depicted as dots in left panel of Fig.~\ref{extremesmall}.

In order to extrapolate the $\tilde \omega_{\rm i}=\tilde\omega_{\rm i}(N\rightarrow 0)$ from the limited data, we use Taylor expansion of the inverse power of the grid point number $N$, and find that these data dots can be fitted with the following function
\bea
\fft{\tilde\omega_{i}(N)}{\tilde\omega_{i}(\infty)} &\sim& 1+ \frac{8166.6}{N} -\frac{7.6951\times 10^8}{N^2} + \frac{5.7386\times 10^{12}}{N^3}
-\frac{1.2714\times 10^{16}}{N^4} + \frac{9.6909\times 10^{18}}{N^5}\,,\nn\\
\tilde\omega_{\rm i}(\infty) &\sim & 2.5246\times 10^{-58}\,.\label{smallextra}
\eea
It can be seen from the solid line in the left panel of Fig.~\ref{extremesmall} that the function fits with the numerical data (dots) quite well for $N$ from 3000 to 10000. The errors at each data point, defined by
\be
\delta_{\rm i} =\fft{|\tilde \omega_{\rm i}(N) - \tilde \omega_{\rm i}|}{\tilde\omega_{\rm i}}\,,
\ee
are depicted in the right panel of Fig.~\ref{extremesmall}. They are all less than 0.3\%. The function \eqref{smallextra} indicates that we need to have $N$ to be a few tens of thousands in order to get $\tilde \omega_{\rm i}(N)$ close enough to be the final convergent value, and it is beyond our computer power. It should be pointed out that $\tilde \omega_{\rm i}(\infty)$ obtained in the above can only be trusted by its magnitude as around $10^{-58}$, not by its precise value, since the curve fitting by a different power of $1/N$ can alter the answer.

It is natural to ask the question, what is the $\tilde\omega_{\rm i}$'s $\tilde \beta$ dependence as $\tilde \beta \rightarrow 0$?  We find that for a given $N\ge 2500$, numerical data accurately indicate that
\be
\tilde\omega_{\rm i}(N, \tilde\beta) \sim \tilde\omega_{\rm i}(N)\tilde \beta^6\,,\qquad \hbox{as}\qquad \tilde \beta\rightarrow 0\,.
\ee
We verify this equation for $\tilde \beta=10^{-3}, 10^{-4}, 10^{-5}, 10^{-7}$, with $N=2500, 3000, 3000$. Together with the extrapolated result \eqref{smallextra}, we have
\be
\tilde\omega_{\rm i} \sim \tilde \omega^0_{\rm i} \tilde \beta^6\,,\qquad\hbox{as}\qquad \tilde\beta \rightarrow 0\,,\qquad \hbox{with}\,,\qquad \tilde \omega^0_{\rm i}\sim 10^{-16}\,.
\ee
Thus we see that when $\tilde \beta\rightarrow 0$, the real and imaginary parts of the dimensionless frequency has $\tilde\beta$ dependence of $\tilde \beta^4$ and $\tilde \beta^6$ respectively.

\subsection{Negative $\tilde\beta$}

As we have discussed in section \ref{sec:theory}, when the coupling constant $\beta$ becomes negative, the dominant energy is violated, but both strong and weak energy still survive. However, the theory is no longer a good effective theory from the point of view of WGC. We therefore do not focus on the negative coupling constant.
Nevertheless, we obtain the ground QBS for $\tilde\beta=-1/2$, by the shooting-target method. The complex frequency is
\be
\tilde\omega= (1-2.49598\times 10^{-7}) + 1.40462\times 10^{-32}\,{\rm i}\,.
\ee
Furthermore, since $\tilde \omega$ is an even function of $\tilde \beta$ as it approach $0^+$, we expect the same result as $\tilde\beta \rightarrow 0^-$, and we verified with a few examples that this is indeed the case.

\section{Conclusions}
\label{sec:conclusion}

One motivation of this work was inspired by the fact that although the extremal RN black hole is superradiantly stable, in its STU supergravity generalization into four different constituents as D-branes and KK monopoles, no matter how small the deviation is, the black hole becomes superradiantly unstable, with both unstable QBS's \cite{Mai:2021yny} and QNM's \cite{Mai:2022thu}. We therefore considered another bigger theory, namely the BI generalization of the Maxwell theory, which has a continuous parameter deviating away from the Maxwell theory. We studied superradiant instability of its charged extremal black hole, by a minimally coupled charged massive scalar perturbation. Our data suggests that the charged extremal black hole in EBI theory are always superradiantly unstable due to the existence of unstable QBS's. Thus the stability of the RN black hole is again a fine-tuned result from the point of view of EBI, as in the case of the STU supergravity model.

Another motivation was inspired by the WGC, which states that quantum correction should make gravity weaker than the gauge interaction; consequently, the extremal RN black hole becomes unstable under the quantum correction and decay into smaller ingredients. The WGC was illustrated to be true for EM gravity extended by the most general four-derivative terms in the context of effective field theory as the leading-order correction \cite{Cheung:2018cwt}. At the linear order of coupling $\beta$, BI action is the subset of the most general corrections considered in \cite{Cheung:2018cwt}, but it satisfies the WGC condition, namely $\Delta S>0$ and $\Delta M<0$ for $\beta >0$. Therefore, it can be used as a toy model to study the stability under quantum correction. The general conclusion appears to be correct: no matter how small the dimensionless coupling $\tilde \beta$ is, the extremal black is superradiantly unstable because of the existence of unstable QBS's. However, it should be pointed out, for reasons not entirely clear, that the imaginary part of the complex frequency is exceedingly small, and hence it would take a long halftime for the instability to take effect.

There is another subtlety in the detail. For small $\tilde \beta$, we find that the dimensionless complex frequency of the unstable QBS's has the $\tilde \beta$ dependence as
\be
\tilde \omega \sim  (1 - 2.46906\times 10^{-6}\, \tilde \beta^4) + 10^{-16} \tilde \beta^6\, {\rm i}\,,\qquad \hbox{as}\qquad \tilde\beta \rightarrow 0\,.
\ee
This implies that we would not have found these unstable QBS's had we restricted ourself to the linear order of $\beta$, as one would in the perturbative effective field theory approach. Our analysis therefore provides another example of instability at the full nonlinear level, despite being stable at the linear level. A further puzzling issue is that despite of the fact that gravity  becomes stronger than the electrostatic force at the long distance for negative $\beta$, the charged extremal black holes still suffer from the superradiant instability. Our results indicate that under full nonlinear effect, large extremal particle-like black holes may break up and decay even when gravity is not the weaker force, presenting a challenge to the WGC statement.

\section*{Acknowlegement}

We are grateful to Yang Huang and Hong-Sheng Zhang for giving us a pedagogical introduction to the spectral method. This work is supported in part by the National Natural Science Foundation of China (NSFC) grants No.~11935009 and No.~12375052.

\end{document}